\documentclass[aps,twocolumn,showpacs,superscriptaddress,pra]{revtex4}
\usepackage{graphicx,amsmath,color}
\usepackage{bm}

\begin{document}

\title{Spin-orbit-coupled Bose-Einstein condensates held under toroidal trap}

\author{Xiao-Fei Zhang}
\affiliation {Department of Engineering Science, University of Electro-Communications, Tokyo 182-8585, Japan}
\affiliation {Key Laboratory of Time and Frequency Primary Standards, National Time Service Center, Chinese Academy of Sciences, Xi'an 710600, China}

\author{Masaya Kato}
\affiliation {Department of Engineering Science, University of Electro-Communications, Tokyo 182-8585, Japan}

\author{Wei Han}
\affiliation {Key Laboratory of Time and Frequency Primary Standards, National Time Service Center, Chinese Academy of Sciences, Xi'an 710600, China}


\author{Shou-Gang Zhang}
\affiliation {Key Laboratory of Time and Frequency Primary Standards, National Time Service Center, Chinese Academy of Sciences, Xi'an 710600, China}

\author{Hiroki Saito}
\affiliation {Department of Engineering Science, University of Electro-Communications, Tokyo 182-8585, Japan}

\date{\today}

\begin{abstract}

We study a quasispin-$1/2$ Bose-Einstein condensate with synthetically generated spin-orbit coupling
in a toroidal trap, and show that the system has a rich variety of ground and metastable states. As
the central hole region increases, i.e., the potential changes from harmonic-like to ring-like, the
condensate exhibits a variety of structures, such as triangular stripes, flower-petal patterns, and
counter-circling states. We also show that the rotating systems have exotic vortex configurations.
In the limit of a quasi-one dimensional ring, the quantum many-body ground
state is obtained, which is found to be the fragmented condensate.
\end{abstract}

\pacs{03.75.Lm, 03.75.Mn, 67.85.Bc, 67.85.Fg}

\maketitle

\section{Introduction}

The engineering of synthetic gauge field and spin-orbit coupling (SOC) in neutral atomic gases has recently
attracted major attention both theoretically and experimentally~\cite{Y.-J. Lin1,Y.-J. Lin2,Y.-J. Lin3,Y.-J. Lin4,
P. Wang,B. M. Anderson,Z. Fu,S. C. Ji,L. Huang,Z. Wu}. In condensed matter physics, the SOC plays an important
role for the emergence of many exotic quantum phenomena~\cite{M. Z. Hasan,X.-L. Qi}. The creation of SOC in
spinor Bose-Einstein condensates (BECs) not only offers us a new platform to simulate the response of charged
particles to external electromagnetic field, but also opens up an entirely new paradigm for studying strong
correlations of quantum many-body systems, which enables quantum simulations of condensed matter phenomena
because of the high controllability of the system~\cite{J. Dalibard,C. J. Wu,N. Goldman,H. Zhai0}.
Very recently, another type of SOC, namely, the spin and orbital-angular-momentum coupling, has also
been proposed~\cite{K. Sun,M. Demarco,C. Qu}.

For a homogeneous spin-orbit (SO) coupled condensate, the mean-field ground state favors
either a plane wave or a striped wave depending on the ratio between inter- and intra-component
interactions~\cite{C. Wang,T. L. Ho}. The presence of the trapping potential modifies this situation
and leads to a rich ground-state physics~\cite{S. Sinha,H. Hu,Y. Zhang,Y. Li,T. Kawakami,B. Ramachandhran,E. Ruokokoski}.
In the presence of a two-dimensional (2D) harmonic trap, a complex phase diagram of Rashba
SO-coupled Bose gases was observed, in which there are two classes of phases and several
subphases~\cite{H. Hu}. In addition, SO coupled BECs subject to rotation have been studied, which
exhibit a rich variety of the ground-state phases and vortex configurations depending on the
strength of SOC and rotation frequency~\cite{X.-Q. Xu1,J. Radic,X.-F. Zhou,A. Aftalion,H. Sakaguchi}.

Bosonic gases loaded in a toroidal trap have attracted considerable interest~\cite{S. Eckel,
A. A. Wood,F. Jendrzejewski,L. Corman,S. Beattie,K. C. Wright,C. Ryu0}, where such a
trapping potential can be realized by a blue-detuned laser beam to make a repulsive potential barrier
in the middle of a harmonic magnetic trap~\cite{C. Ryu}. The toroidal trap provides
us an ideal platform to study fascinating properties of a superfluid,
such as persistent flow~\cite{S. Bargi,M. Abad,A. M. Mateo} and symmetry breaking localization~\cite{Saito,Saito1}.
Now the point is that, in the presence of a toroidal trap, we are inquisitive about whether such a potential can
essentially change the properties of a SO coupled BEC with or without rotation, which is what we attempt to do in this work.

The paper is organized as follows. In Sec. II we formulate the theoretical model describing
the SO coupled BECs held under toroidal trap. Various ground and metastable states generated
by the effects of the SOC and toroidal potential are investigated using the mean-field
theory in Sec. III. The quantum many-body ground state is studied in the limit of quasi-1D
ring in Sec. IV. The main results of the paper are summarized in Sec. V.

\section{Model}

We consider a two-component BEC with Rashba SOC confined in a quasi-2D
toroidal trap on the $x$-$y$ plane.
The second-quantized Hamiltonian of the system is given by
$\hat{\mathcal{H}} =\hat{\mathcal{H}}_0+ \hat{\mathcal{H}}_{\rm int}$, where

\begin{eqnarray}
\hat{\mathcal{H}}_0 \!\! &=& \!\!\int d \textbf{r} \hat{\bm{\psi }}^{\dag}
\biggl[ -\frac{\hbar^2 \nabla^2}{2M}
\!+\! \mathcal{V}_{so} \!+\! V(r) - \Omega \hat{L}_z \biggr] \hat{\bm{\psi}}, \nonumber\\
\hat{\mathcal{H}}_{\rm int}\!\! &=& \!\!\int d \textbf{r}
\biggl( \frac{\textmd{g}_{\uparrow\uparrow}}{2}
\hat\psi_\uparrow^{\dagger 2} \hat\psi_\uparrow^2
+ \frac{\textmd{g}_{\downarrow\downarrow}}{2}
\hat\psi_\downarrow^{\dagger 2} \hat\psi_\downarrow^2
+ \textmd{g}_{\uparrow\downarrow} \hat\psi_\uparrow^\dagger
\hat\psi_\downarrow^\dagger \hat\psi_\downarrow \hat\psi_\uparrow
\biggr),
\nonumber \\
\label{H}
\end{eqnarray}
where $\hat{\bm{\psi}}=(\hat{\psi}_{\uparrow},\hat{\psi}_{\downarrow})^{T}$ denotes the field operator of the atom with pseudospin
state $ \uparrow, \downarrow$, and $M$ is the atomic mass. The Rashba SOC is $\mathcal{V}_{\text{so}}\!= \!-i \kappa
(\sigma_x \partial_x  + \sigma_y \partial_y )$ with $\sigma_{x,y}$ being the Pauli matrices and $\kappa$ is the
strength of the SOC. $\Omega$ is the rotation frequency and $\hat{L}_z$ is the $z$-component of
the orbital angular momentum operator, where we assume that the ``SOC
lasers" are also rotated with the system to simplify the problem~\cite{J. Radic}.
Here we further assume that the two intracomponent interaction parameters
are the same, $\textmd{g}_{\uparrow\uparrow}
\!=\!\textmd{g}_{\downarrow\downarrow}\!\equiv\!\textmd{g}$.
When the quasi-2D system is realized by a tight harmonic potential with
frequency $\omega_z$, the effective interaction parameters are given by
$\textmd{g}\!=\!\sqrt{8 \pi} \hbar^2 a  / (M a_z)$ and
$\textmd{g}_{\uparrow\downarrow}\!=\!
\sqrt{8 \pi} \hbar^2 a_{\uparrow\downarrow}  / (M a_z)$,
where $a$ and $a_{\uparrow\downarrow}$ are the corresponding $s$-wave
scattering lengths and $a_z \!=\! \sqrt{\hbar / (M \omega_z)}$.

The trapping potential considered here is a toroidal trap, which reads
\begin{equation}
V(r) = \frac{1}{2} M \omega_{\bot}^2 r^2 + V_0 e^{-2 r^2 / \sigma_0^2} \label{trap},
\end{equation}
where $\omega_{\bot}$ is the radial trap frequency of the harmonic
potential, $r^2=x^2+y^2$, and $V_0$ and $\sigma_0$ are proportional to the
intensity and beam waist of the optical plug.
The bottom of the potential in Eq.~(\ref{trap}) is located at
\begin{equation}
R = \frac{\sigma_0}{\sqrt{2}} \sqrt{\ln \frac{4 V_0}{M \omega_\perp^2
\sigma_0^2}}.
\end{equation}
Expanding Eq.~(\ref{trap}) around $r = R$ and neglecting the third and
higher orders of $r - R$, we obtain
\begin{equation} \label{omega0}
V_r(r) = \frac{M \omega_0^2}{2} (r-R)^2,
\end{equation}
where $\omega_0 = 2 \omega_\perp R / \sigma_0$ and a constant is omitted.
This approximation is valid, when $\hbar \omega_0$ is much larger than the
characteristic energy of the system.

\section{Mean-field analysis}

\subsection{Ground states}

We implement the mean-field approximation by replacing the field-operators
$\hat{\psi}_{\uparrow,\downarrow}$
with the macroscopic wave functions $\psi_{\uparrow,\downarrow}$
in Eq.~(\ref{H}), which gives the
Gross-Pitaevskii energy functional.
We numerically minimize it by using the imaginary
time evolution method and obtain the ground-state wave function.
We work in dimensionless units by scaling with the appropriate factors of
the harmonic trap energy $\hbar \omega_{\bot}$ and the harmonic trap
length $\sqrt{\hbar/ (M \omega_{\bot})}$.
The trapping potential in Eq.~(\ref{trap}) can be rewritten as
$V_r(r) =r^2 /2 + A e^{- r^2 / l^2}$ with
$l\!=\!\sqrt{M\omega_{\bot} \sigma_0^2/ (2\hbar)}$
and $A \!=\!V_0/ (\hbar \omega_{\bot})$.
The trapping potential in Eq.~(\ref{omega0}) is
$V(r)= \tilde{\omega}_0^2 (r-R)^2 / 2 $ with
$\tilde{\omega}_0 = \omega_0 / \omega_{\bot}$.

\begin{figure}[tbp]
\includegraphics[width=8.5cm,clip]{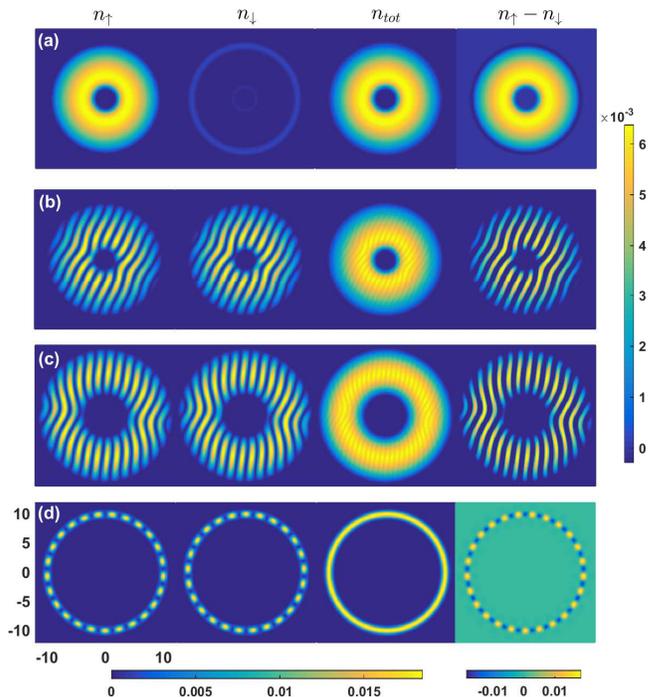}
\caption{(Color online) Examples of ground-state density profiles for the immiscible case. From left to right: densities of
up- and down-components $n_{\uparrow}$ and $n_{\downarrow}$, total density
$n_{\uparrow}+n_{\downarrow}$, and the density difference $n_{\uparrow}-n_{\downarrow}$.
Panels (a) and (b) are for $l=2$, but for $\kappa=1$ and $2$, respectively; while (c) is for $l=5$ and $\kappa=2$.
Other parameters are given as $\textmd{g}=6000$, $\textmd{g}_{\uparrow\downarrow}=8000$, and $A=100$. (d) Example
of density profile of the system in a one-dimensional ring trap of
Eq.~(\ref{omega0}) with $\textmd{g}=100$, $\textmd{g}_{\uparrow\downarrow}=120$,
$\tilde{\omega}_0=\sqrt{10}$, $R=10$, and $\kappa=1$.
}\label{fig1}
\end{figure}

Figure \ref{fig1} shows the ground-state density profiles of the system
for $\textmd{g} < \textmd{g}_{\uparrow\downarrow}$, i.e., the two
components are immiscible.
For small $l$ and $\kappa$, the system exhibits the radial phase
separation, where one component is surrounded by the other one, as shown
in Fig. \ref{fig1}(a). If the SOC is absent and the ratio $\int |\psi_\uparrow|^2 d\textbf{r} /
\int |\psi_\downarrow|^2 d\textbf{r}$ is not fixed, the ground state is
occupied only by one component for the immiscible case.
The two thin rings of $n_\downarrow$ in Fig.~\ref{fig1}(a) is therefore the effect of the
SOC. For strong SOC, the rotational symmetry is broken and the
ground state shows modified stripe phase, as shown in Figs. \ref{fig1}(b) and
\ref{fig1}(c), which are similar to those predicted for homogeneous condensates except for
the central hole region~\cite{C. Wang,S. Sinha}. Near the central hole, the stripe is bent, and it seems that the stripe
tends to be perpendicular to the perimeter of the central hole. In the limit of very tight and narrow
annulus, this tendency becomes significant; the ground-state density distribution of the system shows alternately
arranged stripe pattern along the ring, as shown in Fig.~\ref{fig1}(d).
The number of stripes increases with the strength of the SOC.

\begin{figure}[tbp]
\includegraphics[width=8.5cm,clip]{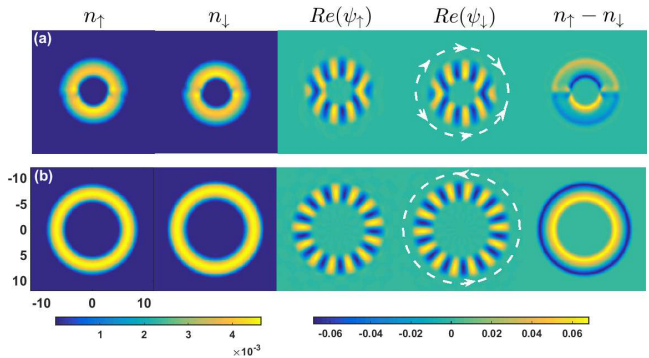}
\caption{(Color online) Typical examples of the ground states for the miscible case. From left to right: densities of
up- and down- components $n_{\uparrow}$ and $n_{\downarrow}$, the real
parts ${\rm Re}(\psi_{\uparrow})$ and ${\rm Re} (\psi_{\downarrow})$, and the density difference $n_{\uparrow}-n_{\downarrow}$.
The parameters are given as $\textmd{g}=1000$, $\textmd{g}_{\uparrow\downarrow}=800$, $A=100$, $\kappa=2$, and $l=2$ in (a) and $l=5$ in (b).
The direction of the mass flow is marked by the dashed-line arrows.
}\label{fig2}
\end{figure}

Figure \ref{fig2} shows the miscible cases with $\textmd{g} > \textmd{g}_{\uparrow\downarrow}$.
When the central hole is small, the ground state is similar to the plane wave~\cite{C. Wang}, where all
the bosons are condensed into a single plane-wave state, and the direction of the plane wave is
chosen in the $x$-$y$ plane breaking the rotational symmetry. As the central hole is increased,
the effect of the toroidal potential emerges; the direction of the flow tends to be azimuthal, as shown in
Fig.~\ref{fig2}(a). As a result, the flow pattern becomes counter-circling.
For a larger $l$ (close to the ring limit), this tendency becomes
significant and the flow becomes one-way circulation, as shown in
Fig.~\ref{fig2}(b). Thus, the direction of the mass flow in
the ground state changes with the shape of the toroidal trap.

\subsection{Variational analysis}

The results in Figs.~\ref{fig1} and \ref{fig2} indicate that the direction
of the wave-number vector tends to be azimuthal in a toroidal trap, and this
tendency is more significant for a tighter toroidal confinement.
To understand this result, we employ the Gaussian variational method.
For simplicity, we assume a potential $V=x^2/2$, and show that the
wave-number vector becomes perpendicular to the confinement, i.e., in the
$y$ direction. For the miscible case, the variational wave functions are assumed to be
\begin{eqnarray}
\psi_\uparrow&=& \frac{1}{\sqrt{2}}\frac{1}{\pi^{1/4}\sigma^{1/2}} e^{-\frac{(x+X)^2}{2\sigma^2}}e^{i \textbf{k} \cdot \textbf{r}},  \nonumber\\
\psi_\downarrow&=& \frac{1}{\sqrt{2}}\frac{-e^{i\phi_{\textbf{k}}}}{\pi^{1/4}\sigma^{1/2}} e^{-\frac{(x-X)^2}{2\sigma^2}}e^{i \textbf{k} \cdot \textbf{r}},
\label{VF}
\end{eqnarray}
where $\sigma$, $X$, and $\textbf{k}$ are variational parameters and
$\phi_{\bf{k}}$ is the angle between $\textbf{k}$ and the $x$ axis.
The shift $X$ in Eq.~(\ref{VF}) is justified by the numerical result shown
in the last column of Fig.~\ref{fig2}, where $|\psi_\uparrow|^2$ and $|\psi_\downarrow|^2$
are shifted in the opposite directions perpendicular to the direction of
confinement. Substituting Eq.~(\ref{VF}) into the Gross-Pitaevskii energy functional,
we obtain the variational energy as
\begin{eqnarray}
E &=& \frac{1}{4\sigma^2} + \frac{k^2}{2} + \frac{\sigma^2}{4} +
\frac{X^2}{2} - \kappa \left( k + \frac{k_y}{k} \frac{X}{\sigma^2}
\right) e^{-\frac{X^2}{\sigma^2}}
\nonumber \\
& & + \frac{1}{4\sqrt{2\pi}\sigma}
\left(\textmd{g} + \textmd{g}_{\uparrow\downarrow}
e^{-\frac{X^2}{\sigma^2}} \right).
\label{E}
\end{eqnarray}
We note that this energy is independent of the direction of $\textbf{k}$,
if $X = 0$.
Assuming $X \ll \sigma$, we find that $k_x = 0$, $k_y \simeq \pm \kappa$,
and $X \simeq \pm \kappa / \sigma^2$ minimize $E$.
The energy of the state with $\textbf{k}$ in the $y$ direction is smaller
than that with $\textbf{k}$ in the $x$ direction by
$\simeq \kappa^2 / (2 \sigma_0^4)$.
Thus, in the miscible case, the ground state tends to have momentum
perpendicular to the direction of confinement.

For the immiscible case, the variational wave function is assumed to be
\begin{eqnarray}
\psi_\uparrow \!\!&=\!\!& \frac{1}{2 \pi^{1/4}\sigma^{1/2}} \left[ e^{-\frac{(x-X)^2}{2\sigma^2}} \! e^{i \textbf{k}
\cdot \bf{r}} + e^{-\frac{(x+X)^2}{2\sigma^2}} \! e^{-i \textbf{k} \cdot \bf{r}} \right], \nonumber\\
\psi_\downarrow\!\!&=\!\!& \frac{-e^{i\phi_{\bf{k}}}}{2 \pi^{1/4}\sigma^{1/2}} \left[ e^{-\frac{(x+X)^2}{2\sigma^2}}
\! e^{i \textbf{k} \cdot \bf{r}} - e^{-\frac{(x-X)^2}{2\sigma^2}} \! e^{-i \textbf{k} \cdot \bf{r}} \right],
\label{VF2}
\end{eqnarray}
which reduces to the usual stripe state for $X = 0$.
With the same procedure as above (see Appendix), we find that the energy
is minimized by $k_x = 0$, $k_y \simeq \pm \kappa$, and $X \simeq \pm
\kappa / \sigma^2$, which are the same as in the miscible case.
Thus, in both miscible and immiscible cases, the direction of the
wave-number vector $\textbf{k}$ tends to be perpendicular to the direction
of confinement, which explains the behaviors in Figs.~\ref{fig1} and
\ref{fig2}, i.e., the wave-number vector tends to be azimuthal in a
toroidal trap.

\subsection{Various metastable states}

\begin{figure}[tbp]
\includegraphics[width=7.5cm,clip]{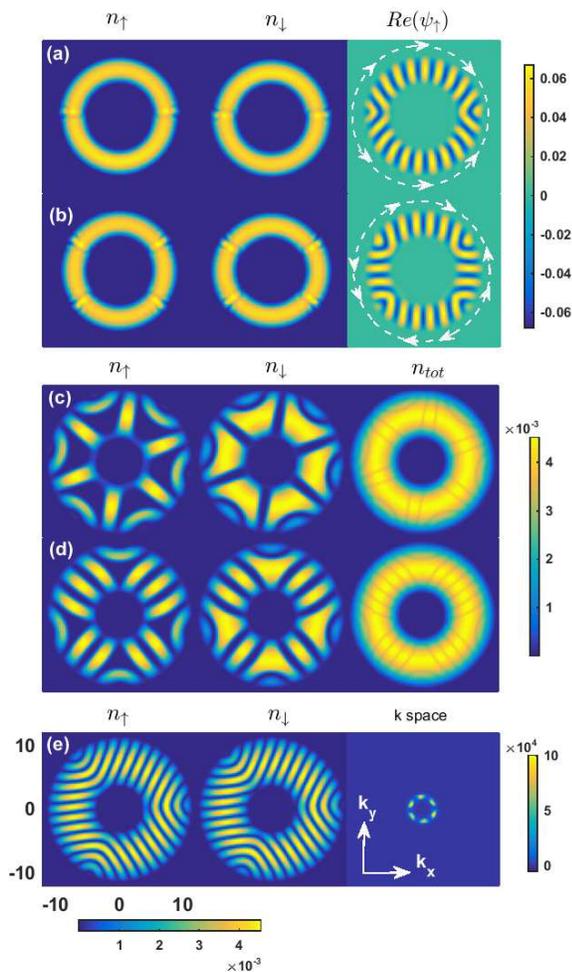}
\caption{(Color online) Typical stationary states of the system.
(a)-(b) Two different types of density and flow patterns for the miscible
case with $\textmd{g}=1000$, $\textmd{g}_{\uparrow\downarrow}=800$, $A=100$, $l=5$, and
$\kappa=3$, where the direction of the mass flow is marked by the dashed-line arrows.
(c)-(d) Two different types of density profiles
for the immiscible case with $\textmd{g}=6000$,
$\textmd{g}_{\uparrow\downarrow}=8000$, $A=100$, $l=5$, and $\kappa=1.25$; while
(e) density and momentum distributions for $\kappa=2$.
}\label{fig3}
\end{figure}

There are various metastable states which have almost the same energy as
the ground state, as shown in Fig.~\ref{fig3}.
To trigger these various pattern formations, we start from different
initial states in the imaginary-time propagation.

Figures~\ref{fig3}(a) and \ref{fig3}(b) show two typical density
distributions and its associated real part of the wave
functions for the miscible case with the same parameter set, where the direction of the mass flow is marked by the dashed-line arrows.
It is clearly seen that different types of the mass flow can be triggered by a proper choice of the initial state.

Figures~\ref{fig3}(c) and~\ref{fig3}(d) show two different types of
stationary states for the immiscible case with the same parameter set.
The density distributions of these metastable states show flower-petal stripe
patterns, which are quite different from the usual stripe phase.
Interestingly, with a further increase in SOC, the stripe shows triangular structure, as shown
in Fig.~\ref{fig3}(e) for $\kappa=2$. To get a deeper physical insight into the origin of this state, it is useful to look at its
momentum distribution, which is presented in the right panel of Fig.~\ref{fig3}(e).
One can see that six maxima at angles $\varphi =\pi/3$ appear and form a ringlike momentum
distribution. This is in a sense reminiscent of the harmonically-trapped case, where
a lattice phase emerges as the ground state~\cite{S. Sinha}. In the present case,
large $l$ leads to a large radius of the toroidal trap and a large central hole,
and the behavior of the system is intermediate between the harmonic potential and
one-dimensional (1D) ring. As a result, the density distribution of the system shows the
triangular stripe phase, where main region of the toroidal trap is occupied by the straight
stripes and their vertices with an equilateral triangle shape by curved stripes.

\subsection{Rotational properties}

\begin{figure}[tbp]
\includegraphics[width=8.0cm,clip]{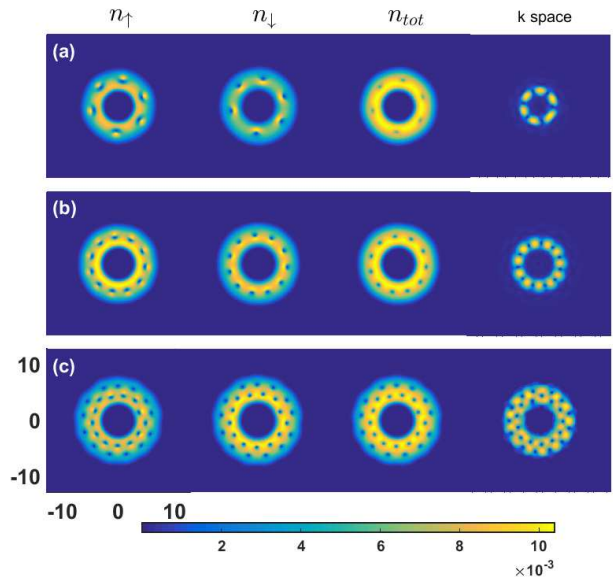}
\caption{(Color online) Ground states of the rotating system for $l=2$,
and for rotation frequencies $\Omega=0.4,0.6$, and $0.8$, corresponding to (a), (b), and (c), respectively. From left to right:
Density of up-component, density of down-component, total density, and $k$-space density of the spin-up component.
Other parameters are given as $\textmd{g}=6000$, $\textmd{g}_{\uparrow\downarrow}=8000$, $\kappa=1.0$, and $A=100$.
}\label{fig4}
\end{figure}

One of the interesting aspects of the toroidal trap is its rotational properties, which manifest themselves
as the presence of quantized vortices and its related vortex lattices~\cite{A. Aftalion1,M. Abad1}. The typical vortex structures
of a harmonically-trapped BECs with SOC and rotation have been well studied, showing various exotic vortex lattices
by varying the strength of SOC and rotation frequency~\cite{X.-Q. Xu1,J. Radic,X.-F. Zhou,A. L. Fetter}.

Our numerical results show that when the rotation frequency $\Omega$ is
small, vortices are distributed on a circle and the number of vortices
along the circle increases with $\Omega$, as shown in Figs.~\ref{fig4}(a)
and \ref{fig4}(b). With an increase in the rotation frequency, we find that not
only the number of vortices along a single circle but also the number of
circles increases, as shown in Fig.~\ref{fig4}(c). For fixed $l$ and $\Omega$, either the number of
vortices on each circle or the number of vortex circles increases with the
strength of SOC, which is easily explained by the fact that the SOC
can help vortex nucleation. Interestingly, the momentum-space densities have similar structures to the
real-space densities, as shown in the last column of Fig.~\ref{fig4}.
There are vortices also in the momentum space and the number of vortices in the
momentum space is equal to that in real space.

\section{Quantum many-body analysis}

We now consider a quasi-1D ring potential, in which $\omega_0$ in
Eq.~(\ref{omega0}) is so large that
$\sqrt{\hbar/ (M \omega_0)} \ll R$
and $\hbar \omega_0$ is much larger than any other energy scales.
In this case, the atoms are confined to $r\simeq R$.
The radial wave function can be approximated by the ground state of
\begin{equation}
\left[- \frac{1}{2} \left(\frac{\partial^2}
{\partial r^2}+ \frac{1}{r} \frac{\partial}{\partial r} \right) + V_r(r)
\right] f(r) =E_0 f(r),
\end{equation}
where $\int_0^\infty f^2(r) r dr = 1$.
One can show $\int_0^\infty f^2(r) dr \simeq 1 / R$ and $\int_0^\infty
f(r) f'(r) r dr \simeq -1 / (2R)$.
Dividing the field operator into the radial and azimuthal parts as
$\hat\psi_{\uparrow, \downarrow}(\textbf{r}) = f(r)
\hat\phi_{\uparrow, \downarrow}(\theta)$,
the Hamiltonian $\hat{\mathcal{H}}$ in Eq.~(\ref{H}) reduces to
\begin{eqnarray}
& & \hat{\mathcal{H}} = \int_0^{2\pi} d\theta \biggl\{
\hat{\bm{\phi}}^\dagger \left(E_0 - \frac{1}{2R^2}
\frac{\partial^2}{\partial r^2} + i \Omega \frac{\partial}{\partial \theta} \right) \hat{\bm{\phi}}
\nonumber \\
& & + \frac{i\kappa}{R} \left[ e^{-i\theta} \hat\phi_\uparrow^\dagger
\left( \frac{1}{2} + i \frac{\partial}{\partial \theta} \right)
\hat\phi_\downarrow + e^{i\theta} \hat\phi_\downarrow^\dagger \left(
\frac{1}{2} - i \frac{\partial}{\partial \theta} \right) \hat\phi_\uparrow
\right]
\nonumber \\
& & + \frac{\textmd{g} C}{2} (\hat\phi_\uparrow^{\dagger 2}
\hat\phi_\uparrow^2
+ \hat\phi_\downarrow^{\dagger 2} \hat\phi_\downarrow^2)
+ \textmd{g}_{\uparrow\downarrow} C \hat\phi_\uparrow^\dagger
\hat\phi_\downarrow^\dagger \hat\phi_\downarrow \hat\phi_\uparrow
\biggr\},
\label{Hphi}
\end{eqnarray}
where $\hat{\bm{\phi}} = (\hat{\phi}_\uparrow, \hat{\phi}_\downarrow)^T$
and $C = \int_0^\infty f^4(r) rdr$. Since the total number of atoms is fixed, we neglect the term of $E_0$ in
the following.
We expand the field operator as
\begin{equation} \label{phi}
\hat{\phi}_{\uparrow,\downarrow}(\theta) = \sum_{n}
\hat{a}_{\uparrow,\downarrow n} \frac{e^{i n\theta}}{\sqrt{2\pi}},
\end{equation}
where $\hat{a}_{\uparrow,\downarrow n}$ satisfies the bosonic commutation
relations.
Substituting Eq.~(\ref{phi}) into the one-body part of Eq.~(\ref{Hphi}),
we obtain
\begin{eqnarray}
\hat{\mathcal{H}}_0 & = & \sum_n \biggl[
\frac{n^2}{2R^2} \left( \hat{a}_{\uparrow n}^{\dag}
\hat{a}_{\uparrow n}+ \hat{a}_{\downarrow n}^{\dag}
\hat{a}_{\downarrow n } \right) \nonumber\\
& & + \frac{i\kappa}{R} m
\left( \hat{a}_{\downarrow n+1 }^{\dag} \hat{a}_{\uparrow n }
- \hat{a}_{\uparrow n }^{\dag}\hat{a}_{\downarrow n+1 } \right) \biggr],
\label{HR}
\end{eqnarray}
where $m = n + 1 / 2$, and we set $\Omega=0$ for simplicity.
Using linear transformation as
\begin{eqnarray}
\hat{\alpha}_m^- & = &
\hat{a}_{\uparrow n} \cos \alpha + i \hat{a}_{\downarrow n+1} \sin \alpha,
\nonumber \\
\hat{\alpha}_m^+ & = &
\hat{a}_{\uparrow n} \sin \alpha - i \hat{a}_{\downarrow n+1} \cos \alpha,
\end{eqnarray}
with $\tan 2\alpha = 2\kappa R$, Eq.~(\ref{HR}) is diagonalized as
\begin{equation}
\hat{\mathcal{H}}_0 = \sum_m \left( E_m^+ \hat{\alpha}_m^{+\dag}
\hat{\alpha}_m^+ + E_m^- \hat{\alpha}_m^{-\dag} \hat{\alpha}_m^- \right),
\end{equation}
where
\begin{equation}
\label{Em}
E_m^{\pm} = \frac{1}{2R^2} \left( m^2 + \frac{1}{4} \pm m \cos 2\alpha
\right) \pm \frac{\kappa}{R} m \sin 2\alpha.
\end{equation}
These energies are two-fold degenerate because of $E_m^- = E_{-m}^+$.

We consider the quantum many-body state that minimizes
$\langle \hat{\mathcal{H}}_0 \rangle$.
The solution of $\partial E_m^{\pm} / \partial m = 0$ is given by
$m = \mp \sqrt{1 + (2\kappa R)^2} / 2$.
Let $\mp m_0$ be the half integers that are closest to
$\mp \sqrt{1 + (2\kappa R)^2} / 2$ and minimize $E_m^\pm$, which we define
$E_{\rm min} = E_{\mp m_0}^\pm$.
The $N$-particle states that minimize
$\langle \hat{\mathcal{H}}_0 \rangle$ are then
\begin{equation}
| N-p, p \rangle  = \frac{(\hat{\alpha}_{-m_0}^{+\dag})^{N-p}
(\hat{\alpha}_{m_0}^{- \dag})^p}{\sqrt{(N-p)!} \sqrt{p!}} |0 \rangle,
\label{mbs}
\end{equation}
where $p=0,1,\cdots,N$.
These states satisfy $\hat{\mathcal{H}}_0 | N-p, p \rangle
= N E_{\rm min} | N-p, p \rangle$, and hence $(N+1)$-fold degenerate.
When $\kappa R \gg 1$, we find $\alpha \simeq \pi / 4$, $m_0 \simeq \kappa
R$, and $E_{\rm min} \simeq -\kappa^2 / 2$.
In this case, the wave number along the ring is
$\simeq m_0 / R \simeq \kappa$.
When $\kappa R < \sqrt{3} / 2$, we find $\alpha < \pi / 6$, $m_0 = 1/2$,
and $n = 0$, and the trivial state in which all the atoms are at rest is
recovered. We therefore consider the case of $\kappa R > \sqrt{3} / 2$ in the
following.

Next we examine the interaction energy. We assume that the interaction energy is
much smaller than the kinetic and SOC energies, and that the ground state is
spanned by the states in Eq.~(\ref{mbs}). The expectation value of the interaction
part of Eq.~(\ref{Hphi}) with respect to the many-body state in Eq.~(\ref{mbs})
is calculated to be
\begin{eqnarray}
& & \langle N-p, p| \hat{\mathcal{H}}_{\rm int} |N-p, p \rangle
\nonumber \\
& = & \frac{C}{16\pi} \left[ 3\textmd{g} + \textmd{g}_{\uparrow\downarrow}
+ (\textmd{g} - \textmd{g}_{\uparrow\downarrow}) \cos 4\alpha \right]
N (N - 1)
\nonumber \\
& & - \frac{C}{8\pi} (\textmd{g} - \textmd{g}_{\uparrow\downarrow})
(1 + 3 \cos 4\alpha) p (N - p).
\label{Hint}
\end{eqnarray}
The matrix element
$\langle N-p, p| \hat{\mathcal{H}}_{\rm int} |N-p', p' \rangle$
vanishes for $p \neq p'$.
In the last line of Eq.~(\ref{Hint}), $1 + 3 \cos 4\alpha$ is negative.
Therefore, the interaction energy is minimized by $p=0$ or $p=N$ for
$\textmd{g} > \textmd{g}_{\uparrow\downarrow}$ and by $p=N/2$ for
$\textmd{g} < \textmd{g}_{\uparrow\downarrow}$, which correspond to the
plane-wave and stripe phases, respectively.
We note that the ground state for $\textmd{g} <
\textmd{g}_{\uparrow\downarrow}$ is $|N/2,N/2\rangle$, which is the
fragmented BEC~\cite{S.-W. Song}.

\section{Conclusions}

We have investigated the effect of a toroidal trap on a SO coupled BECs with or without
rotation. In contrast to the case of the harmonic trap, the interplay between the SOC
and toroidal trap can result in a rich variety of ground and metastable states, such
as triangular stripes, flower-petal patterns, and counter-circling states. We found that
the condensate in a toroidal trap tends to have wave-number vectors in the azimuthal
direction, which becomes more significant for a tighter toroidal trap. This finding can
be explained by the Gaussian variational analysis. In the presence of rotation, the system
is found to exhibit exotic ground-state vortex configurations by varying the strength of SOC
and rotation frequency, and the number of vortices is the same as that in the momentum space.
We have also formulated a quantum many-body problem for the quasi-1D ring trap for both miscible
and immiscible cases. In the latter case, the ground state of the system is found to be a
fragmented condensate. Owing to the recent developments in the experimental implementation of SOC
and the high degrees of control over most of the system parameters, the various states found
in this work may be observed in current experiments.

\begin{acknowledgments}

This work was supported by JSPS KAKENHI Grant Numbers JP16K05505, JP26400414, and JP25103007, by the NSFC under grants
Nos. 61127901, and 11547126, by the key project fund of the CAS for the ``Western Light'' Talent Cultivation Plan under
grant No. 2012ZD02, and by the Youth Innovation Promotion Association of CAS under grant No. 2015334.

X.-F. Zhang and M. Kato contributed equally to this work.

\end{acknowledgments}

\appendix*

\section{Variational analysis in the immiscible case}

Substituting Eq.~(\ref{VF2}) into the Gross-Pitaevskii energy functional
and integrating with respect to $x$, we obtain
\begin{eqnarray}
E_{\rm kinetic} & = & \frac{1}{4\sigma^2} + \frac{k^2}{2}, \\
E_{\rm potential} & = & \frac{\sigma^2}{4} + \frac{X^2}{2}, \\
E_{\rm soc} & = & -\kappa \left( k + \frac{k_y}{k} \frac{X}{\sigma^2}
\right) e^{-\frac{X^2}{\sigma^2}} \nonumber \\
& & + i e^{-k_x^2 \sigma^2} k_y \cos(2 k_x X + \theta) \cos 2 k_y y,
\nonumber \\
\label{Esoc}
\\
E_{\rm interaction} & = & \frac{1}{8 \sqrt{2\pi} \sigma}
\Bigl[ g(1 + 2e^{-\frac{2X^2}{\sigma^2}}) + g_{\uparrow\downarrow}
\nonumber \\
& & + (g_{\uparrow\downarrow} - g)
e^{-2 k_x^2 \sigma^2 - \frac{2X^2}{\sigma^2}} \cos 4k_y y \Bigr],
\label{Eint}
\end{eqnarray}
where $E_{\rm kinetic}$ and $E_{\rm potential}$ are the same as those in
the miscible case in Eq.~(\ref{E}). The second line of Eq.~(\ref{Esoc})
vanishes by integrating with respect
to $y$. Similarly, the second line of Eq.~(\ref{Eint}) also vanishes for $k_y \neq
0$. In this case, assuming $X \ll \sigma$, one finds that $k_x = 0$, $k_y
\simeq \pm \kappa$, and $X \simeq \pm \kappa / \sigma^2$ minimize the
total energy. The energy lowered by the shift $X$ is $\simeq \kappa^2 / (2 \sigma^4)$.
When the immiscible interaction energy dominates the soc energy, the
second line of Eq.~(\ref{Eint}) is minimized by $k_x = k_y = 0$ and $X =
0$, which corresponds to the state shown in Fig.~\ref{fig1}(a).

\end{document}